\documentstyle[12pt]{article}
\renewcommand{\theequation}{\thesection.\arabic{equation}}
\renewcommand{\thesubsection}{\thesection.\arabic{subsection}}

\newlength{\extraspace}
\newlength{\extraspaces}
\newcounter{dummy}

\newcommand{\be}{\begin{equation}
\addtolength{\abovedisplayskip}{\extraspaces}
\addtolength{\belowdisplayskip}{\extraspaces}
\addtolength{\abovedisplayshortskip}{\extraspace}
\addtolength{\belowdisplayshortskip}{\extraspace}}
\newcommand{\ee}{\end{equation}}

\newcommand{\ba}{\begin{eqnarray}
\addtolength{\abovedisplayskip}{\extraspaces}
\addtolength{\belowdisplayskip}{\extraspaces}
\addtolength{\abovedisplayshortskip}{\extraspace}
\addtolength{\belowdisplayshortskip}{\extraspace}}
\newcommand{\ea}{\end{eqnarray}}

\newcommand{\baa}{
\addtocounter{equation}{1}
\setcounter{dummy}{\value{equation}}
\setcounter{equation}{0}
\renewcommand{\theequation}{\thesection.\arabic{dummy}\alph{equation}}
\begin{eqnarray}
\addtolength{\abovedisplayskip}{\extraspaces}
\addtolength{\belowdisplayskip}{\extraspaces}
\addtolength{\abovedisplayshortskip}{\extraspace}
\addtolength{\belowdisplayshortskip}{\extraspace}}
\newcommand{\eaa}{
\end{eqnarray}
\setcounter{equation}{\value{dummy}}
\renewcommand{\theequation}{\thesection.\arabic{equation}}}

\newcommand{\ban}{\begin{eqnarray*}
\addtolength{\abovedisplayskip}{\extraspaces}
\addtolength{\belowdisplayskip}{\extraspaces}
\addtolength{\abovedisplayshortskip}{\extraspace}
\addtolength{\belowdisplayshortskip}{\extraspace}}
\newcommand{\ean}{\end{eqnarray*}}

\newcommand{\newsection}[1]{
\vspace{20mm}
\pagebreak[3]
\addtocounter{section}{1}
\setcounter{equation}{0}
\setcounter{subsection}{0}
\setcounter{footnote}{0}
\begin{flushleft}
{\large\bf \thesection. #1}
\end{flushleft}
\nopagebreak
\smallskip
\nopagebreak}

\newcommand{\newsubsection}[1]{
\vspace{12mm}
\pagebreak[3]

\addtocounter{subsection}{1}
\addcontentsline{toc}{subsection}{\protect
\numberline{\arabic{section}.\arabic{subsection}}{#1}}
\noindent{\sc \thesubsection. #1}
\nopagebreak
\vspace{2mm}
\nopagebreak}

\newcommand{\ie}{{\it i.e.\ }}

\newcommand{\hf}{{\textstyle{1\over 2}}}

\newcommand{\is}{\! & \! = \! & \!}

\newcommand{\nonu}{\nonumber \\[1.5mm]}

\newcommand{\ra}{\rightarrow}

\newcommand{\half}{{\textstyle{1\over 2}}}

\newcommand{\G}{{\mbox{\footnotesize $G$}}}

\newcommand{\Omeg}{{\mbox{\footnotesize $\Omega$}}}

\newcommand{\del}{\partial}

\renewcommand{\footnotesize}{\small}

\begin{document}
\addtolength{\baselineskip}{.12mm}
\input epsf

\thispagestyle{empty}

\vspace{-1cm}

\begin{flushright}
{\sc CERN-TH}.7469/94\\
{\sc PUPT}-1504\\
January 1995
\end{flushright}

\begin{center}
{\large\sc{Black Hole Horizons 
and Complementarity.}}\\[8mm]
{\sc Youngjai Kiem,$\ $Herman Verlinde${}^*$}\\[3mm]
{\it Joseph Henry Laboratories\\[1.5mm]
Princeton University, Princeton, NJ 08544 \\[2mm]
and \\[2mm]
${}^*$  Institute for Theoretical Physics\\[1.5mm]
University of Amsterdam\\[1.5mm]
Valckenierstraat 65, 1018 XE Amsterdam}\\[3mm]
{ and}\\[.3cm]
{\sc Erik Verlinde}\\[3mm]
{\it TH-Division, CERN\\[1.5mm]
CH-1211 Geneva 23\\[2mm]
and\\[2mm]
Institute for Theoretical Physics\\[1.5mm]
University of Utrecht\\[1.5mm]
P.O. BOX 80.006, 3508 TA Utrecht}\\[11mm]

{\sc Abstract}
\end{center}

\vspace{-1mm}
\noindent
We investigate the effect of gravitational back-reaction
on the black hole evaporation process. The standard
derivation of Hawking radiation is re-examined and extended by
including gravitational interactions between the
infalling matter and the outgoing radiation. We find that these interactions
lead to substantial effects. In particular, as seen by an
outside observer, they lead to a fast growing uncertainty
in the position of the infalling matter as it approaches the
horizon. We argue that this result supports the idea of black hole
complementarity, which states that, in the description of the
black hole system appropriate to outside observers, the region behind
the horizon does not establish itself as a classical region
of space-time. We also give a new formulation of this complementarity
principle, which does not make any specific reference to the
location of the black hole horizon.

\newpage

\newsection{Introduction.}

\noindent
Ever since the discovery of black hole evaporation \cite{hawking}
there has been a continuing debate on the relevance of the
gravitational back-reaction to the final quantum state of the radiation.
The  central question is whether back-reaction effects could, even
in principle, bring out the information about the initial
quantum state of the matter that has formed the black hole.  The answer
to this question would be no, essentially by assumption, if one accepts
that the state of the radiation is reliably computed using
free propagation of quantum fields on a fixed classical background
geometry, and that
the gravitational effect of the quantum radiation is accurately described
via an adiabatic change of the background geometry and the mass of
the black hole.  According to this scenario, which has in particular
been advocated by Hawking,\footnote{For more recent explanations of
this point of view, see \cite{steveandy}.} strong
gravitational effects take place too late or too far behind
the horizon to be able to bring out the initial information.

An opposite point of view has been put forward
by 't Hooft \cite{thooft}, who pointed out that from the
perspective of the outside observer, strong gravitational interactions
take place near the horizon between the infalling matter and the out-going
virtual particles describing the Hawking radiation.
He argued that this interaction could
drastically change the standard semi-classical picture of the evaporation
process, and in particular may give rise to a complementarity between
the physical world of the infalling observer and that of the outside observer.
Indeed, for the infalling observer the horizon represents a smooth region of
space-time, but the Hawking particles are not measurable to him.
To the outside observer detecting the Hawking radiation,
on the other hand,  the horizon becomes a strongly interacting region, while
the black hole behind it never seems to establish itself as a classical
part of space-time. According to this alternative physical picture,
which recently has also been advocated by several other authors
\cite{thooftetal,susskind,unpub,englertetal,2dfluct},
there is no longer any a priori reason
why quantum coherence should be destroyed during the evaporation process.

To investigate the possibility of this second scenario,
we will in this paper re-examine the derivation of Hawking
radiation and present a new procedure for studying
the on-set of gravitational back-reaction effects
on the radiation spectrum. Specifically, we will study
the propagation of a quantum state from an initial Cauchy surface
$\Sigma_{initial}$ located some finite distance outside the event horizon
to a final Cauchy surface $\Sigma_{final}$ located at a later time and
much closer to the event horizon (see fig 1.).
Both surfaces are space-like everywhere and can roughly be thought of as
constant-time slices as seen by an outside observer. Eventually, we will
be interested in the time evolution of the state on $\Sigma_{final}$,
as the outside part of it starts to contain more and more of the
outgoing thermal radiation.

Since both Cauchy surfaces are far away from the black hole singularity,
it would at first sight appear to be sufficient to use free field theory
in a fixed background to relate the physical observations made on each of
these surfaces. We will find however that this naive expectation is incorrect.
Instead we will show that gravitational interactions, that take
place between the modes as they propagate from the initial to final Cauchy
surface, become increasingly significant as the time on $\Sigma_{final}$
progresses. These interactions are associated with two types of collisions,
namely between the infalling and virtual out-going particles near the horizon,
and secondly between the out-going virtual particles themselves.
The interaction regions of these two types of virtual processes are
schematically indicated in fig 1a.
In this paper we will mostly be concerned with the first type of
interactions.

\begin{center}
\leavevmode
\epsfysize=7cm
\epsfbox{figu1.eps}
\end{center}
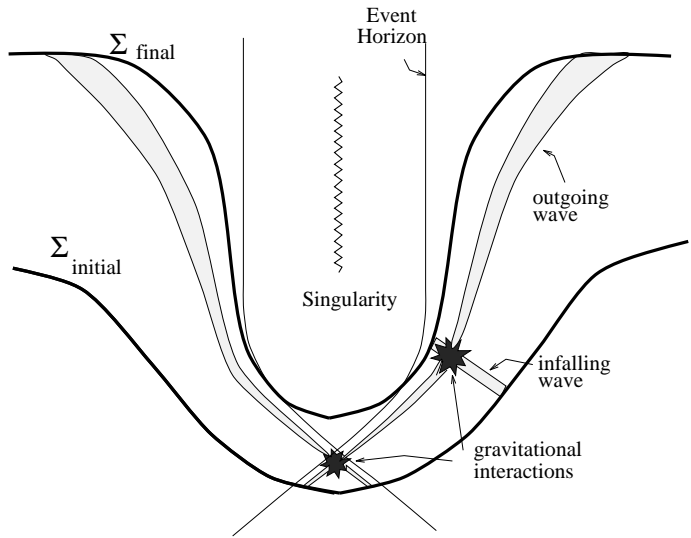
{\small Fig 1a. In this paper we will study the effect of back-reaction on the
propagation of quantum state in a black hole formation geometry. We will
find that even for regular initial data on the initial Cauchy surface
$\Sigma_{initial}$, the virtual gravitational interaction between
the in- and outgoing modes,
as well as among the outgoing modes themselves, will become increasingly
important with time on the final Cauchy surface $\Sigma_{final}$.}

Typically, the in- and outgoing particles involved in these processes
propagate through the horizon at different angular directions.
Hence, while the center of mass energies can be huge, the momentum
transfer during these virtual
collisions is typically small. In this special
kinematical limit, there exist a rather large range of collision
energies for which the quantum gravitational interaction between the
particles is well-controlled and
can be described by means of semi-classical techniques \cite{planckscat}.

This paper is organized as follows. To set up some notation,
we briefly summarize in section 2 the relevant formulas describing
free field propagation in a fixed black hole background.
In section 3 we write the classical equations describing the
gravitational interaction between in and out-going particles near the
horizon. In section 4  we present a new procedure for including the
quantum mechanical effects of the back-reaction. Starting from the
original formulas of Hawking, we will include a small quantum contribution
to the infalling matter that creates the black hole, and compute
its effect on the relation between the $in$ and $out$-modes.
Finally we will then consider the effect of these interactions
on the calculation of the out-going state.

But first we present a somewhat more concrete form of the
complementarity principle put forward in \cite{thooftetal,susskind,unpub}.
Whether this principle is indeed dynamically realized
depends on the yet unknown details of Planck scale physics,  although the
calculations of this paper provide some supporting
evidence.
An important advantage of our formulation is that it does not need to
make any explicit reference to the location of the black hole
horizon, and does not strictly rely on the assumed existence of a
black hole $S$-matrix.

\newcommand{\omegaa}{\omega}

\newsubsection{The complementarity principle.}

Complementarity is a fundamental aspect of quantum mechanics, with
familiar manifestations such as particle-wave duality and the
uncertainty relations between momentum and position operators.
It arises from the fact that a single quantum state occupies
a finite volume of the classical phase space.
This basic property of quantum mechanics also applies
to a scalar field $\phi$ propagating on a dynamical black hole
geometry.  The back-reaction of a classical $\phi$ configuration
can change the background geometry, and thus quantum states are
in principle supported on different classical geometries.
However, we can usually ignore this fact,
because we can in most situations safely truncate the Hilbert
space to a subspace in which quantum gravitational
interactions are guaranteed to be small, e.g. by restricting all modes of
$\phi$ to frequencies sufficiently smaller than the Planck frequency.

Concretely, we can imagine introducing some
position dependent cut-off scale $\epsilon(x)$ on the Cauchy
surface $\Sigma$. We can then represent the corresponding
truncated Hilbert space on $\Sigma$ by means of all states
supported on field configurations with
wavelengths larger than this cut-off scale.
For normal, regular Cauchy surfaces it is reasonable
to expect that a constant cut-off $\epsilon$ of the order
of the Planck length will still be sufficient to
eliminate all strong coupling gravitational effects.
However, in case the Cauchy surface contains
different regions that are related via large relative boosts
(like $\Sigma_{final}$ in fig. 1), it is conceivable that
a much stronger restriction of the free field Hilbert space
may be needed to achieve a reliable semi-classical description.

To address this question, let us first formulate
in a somewhat more precise way the criterion
we would like to impose on the cut-off length scale
$\epsilon(x)$. A first key point is that
to a given cut off $\epsilon(x)$ we can associate a corresponding
size of stress-energy fluctuations. These stress-energy fluctuations
are a necessary consequence of the presence of all modes of the second
quantized scalar field up to the cut-off scale. Their typical size
is determined by the behavior of the regulated expectation value
$\langle T_{\mu\nu}(x)^2\rangle_\epsilon$, and from free field
theory we deduce that the quantum fluctuations of $T_{\mu\nu}$
typically grow like $\epsilon(x)^{-4}$ as the cut-off gets smaller.

Via the Einstein equations, this will result in correspondingly
large quantum fluctuations in the local background geometry
on the Cauchy surface $\Sigma$.
For a given cut-off and Cauchy surface $\Sigma$, one can in
principle calculate these by integrating the cumulative
gravitational effect of the local stress-energy fluctuations.
It is clear that when these geometry fluctuations become too
large, relative to the cut-off
scale, the semi-classical description of the Hilbert space as the space
of states on a given Cauchy slice breaks down. We will call a cut-off
violating this bound {\it super-critical}, and we will call it
{\it sub-critical} or {\it semi-classical} if the fluctuations of
the geometry are controlled in this sense. It is clear from the
above discussion that semi-classical cut-offs have a minimal size.

The question arises, however, how such a semi-classical
truncation of the Hilbert space must be interpreted at a
fundamental level?  On the one hand,  any short distance cut-off
would appear to constitute an unacceptable violation of Lorentz
invariance (= frame-independence), since different observers will
in general be inclined to truncate the Hilbert space in different ways.
On the other hand, there is no correspondence principle that tells us
that the Hilbert space must extend into this super-critical regime.
On the contrary: if we would assume it does extend into the
super-critical regime, we would need to explain why there are
no large space-time fluctuations at scales much larger than
the Planck scale.

Thus it seems we are faced with a dilemma: apparently we must either
give up strict Lorentz invariance, or find a way to deal with a Hilbert
space supported on field configurations with arbitrarily high frequencies.
A way out of this dilemma is provided by the black hole
complementarity principle proposed in \cite{thooft,unpub,susskind}.

We now propose a new formulation of this principle, which we name
{\it space-time} complementarity,  because its formulation and possible
consequences are in principle not restricted to the black hole context.
The spirit in which this principle is meant is as a proposal for
a reasonable effective description of some underlying, consistent theory
of quantum gravity, such as the one provided by string theory.\footnote{
L. Susskind has also advocated that a complementarity principle of the
type formulated in this section may be realized in string theory.
There indeed exist several indications that, compared to local field theory,
string theory has drastically fewer degrees of freedom at short distances.}
Clearly, however, a much more detailed
knowledge of this underlying fundamental theory is necessary to
verify and quantify this effective description.

\medskip

\parbox{15cm}{Space-time complementarity:
{\it A variable cut-off scale $\epsilon(x)$ on a Cauchy surface $\Sigma$
provides a permissible semi-classical description of the second
quantized Hilbert space, only when the quantum fluctuations of the
local background geometry induced by the corresponding stress-energy
fluctuations do not
exceed the cut-off scale itself. All critical cut-off scales that
saturate this requirement provide complete, complementary descriptions of
the Hilbert space.}}\\

The key step here is that, while different observers may under
certain circumstances be inclined to use very different cut-offs, and thus
very different bases of observables for doing measurements, the
Hilbert spaces spanned by these different bases is assumed to be the
same! The consequences of this assumption are particularly striking in
a situation with
two different observers whose reference frames are related by very
large red- or blue-shifts, such as the infalling and outside observer
on a black hole background.

To illustrate this, let us consider the simultaneous measurement of
an $outside$ observable ${\cal O}_{\lambda_{out}}$, supported
on free field configurations of typical wavelength $\lambda_{out}$,
and an $inside$ observable ${\cal O}_{\lambda_{in}}$
near or just behind the horizon,  of typical
wavelength $\lambda_{in}$. (See fig. 1b.)
One would expect that such a simultaneous measurement should
indeed be possible, since the two observables on $\Sigma_{final}$
are space-like separated.

\begin{center}
\leavevmode
\epsfysize=7cm
\epsfbox{figu1b.eps}
\end{center}
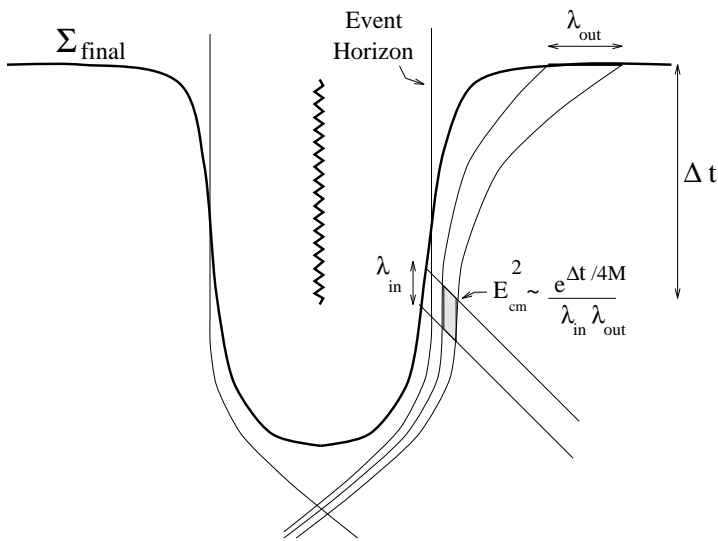
{\small Fig 1b. This figure shows two space-like separated observables defined
on $\Sigma_{final}$ of typical wavelength $\lambda_{in}$
and $\lambda_{out}$. The field modes associated with these
observables have collided in the past with a center of mass
energy that grows as $\exp (\Delta t/8M)$. The proposed complementarity
principle states that observables for which this collision
energy exceeds some (possibly macroscopic)
critical value do not simultaneously exist as mutually
commuting operators.}

\medskip

To accurately compute transition amplitudes involving these operators,
however, we must consider their past history. In first instance,
we can try to compute this past history using the free field
propagation of the modes contained in these observables.
We then discover that the past history of these observables in fact
contains an ultra-high energy collision very close to the horizon,
with a center of mass energy that grows exponentially in the out-going time!
While it is true that this interaction takes place between virtual
instead of real particle excitations, the absurd magnitude of
the collision energy nevertheless indicates that the classical
geometry is no longer the appropriate setting
for considering the simultaneous measurement by these two
observables.

In the following sections we will make this intuition more
explicit, by showing that the relevant transition amplitudes or correlation
functions indeed always involve very large
stress-energy fluctuations, that collide near the horizon (see fig. 1b).
The fact that this stress-energy is associated with virtual particles
implies that it is of purely negative frequency in a local inertial frame.
Although this means that the $in$-$in$ expectation value of the
stress-energy remains small, we will show that its large quantum
fluctuations still lead to non-trivial quantum gravitational effects
that affect the calculation of the out-going state.

Moreover, at a more fundamental level, this observation suggests
that something could actually be wrong in the assumption that both
these observables must be simultaneously present in the Hilbert space
as two mutually commuting operators. Again, as before,
there is no longer any
correspondence principle that tells us that this has to be the case.
Instead it seems to us that
a strong case can be made for the opposite assumption, which provides
the basis for the second formulation of the complementarity principle.

\medskip

\parbox{15cm}{Space-time complementarity (kinematical):
{\it Different microscopic observables that are space-like separated
on a Cauchy surface $\Sigma$, but have support on matter
field configurations that, when propagated back in time,
have collided with macroscopically large center of mass energies, are
not simultaneously contained as commuting operators in
the physical Hilbert space. Instead such operators are
complementary.}}\\

\noindent
Let us end this section with a few short comments:

It is important to emphasize that
in both formulations of the complementarity assumption,
no specific reference is made to the presence of the black hole
horizon, nor to its location. The horizon region is therefore not
considered differently from any other region of space-time.

Nevertheless,  the complementarity assumption
will have drastic consequences for the horizon region as seen by an
observer who stays outside the black hole.
This is particularly evident from the second formulation of the
complementarity principle, since it immediately implies that,
to an outside observer, the part of the Hilbert
space that is associated to the region near or behind the horizon
must be much, much smaller than would follow from free field theory.
It is clear that this will have important consequences
for the computation of quantities like the black hole entropy (cf
ref. \cite{thooft2}).

While the kinematical and dynamical formulation are very similar
in spirit, it is not immediately obvious that they are equivalent.
To establish this equivalence, one would have to show that
the above kinematical complementarity restriction is dynamically
implied by the first principle. In other words, one would
need to show that the simultaneous existence of
operators like the in- and outside operators in fig 1b
inevitably results in macroscopically large space-time fluctuations,
thereby violating the restriction formulated in the first form
of the complementarity principle.
The aim of the following sections is to present evidence that this is
indeed the case.

\pagebreak

\newsection{Scalar fields in a black hole geometry}

In this section we consider the propagation of a free
scalar field on the time-dependent geometry  of a black hole that is
being formed by gravitational collapse of a spherical body of matter.
Here we will ignore back-reaction.
The metric outside the collapsing body is given by the
Schwarzschild metric for a black hole with constant mass $M$,
while inside the matter distribution the metric is assumed to be regular,
and not very different from Minkowski space. It is convenient to introduce
the advanced and retarded time coordinates $v$ and $u$ which in
the Schwarzschild region are
defined by
\be
 u = t- r^* \qquad\quad v = t+r^*
\label{uvcone}
\ee
with
$$
r^* = r + 2M \log \Bigl( \frac{r}{2M} -1 \Bigr) - 2M,
$$
where $r$ and $t$ represent the
usual Schwarzschild radial and time coordinates. In terms of the Kruskal
coordinates $x^+ \! = \! e^{v/ 4M}$ and $ x^- \! = \! -e^{-u/4M}$
the metric outside the collapsing matter is given by
\be
\label{kruskal}
ds^2 = - {32 M^3\over r} e^{-r/2M} dx^+ dx^- +
r^2 d^2 \Omeg,
\ee
where $d^2 \Omeg = d\theta^2 + \sin^2\theta d\varphi^2$ is the line-element
on the sphere.

Now consider the classical propagation of a scalar field $\phi$ on
this geometry. Following \cite{hawking} we want to determine the
relation between a
given out-going wave and the corresponding initial wave that
is obtained by propagating the former backwards in time
through the collapsing matter.
We may concentrate our discussion to the region close to the
horizon, where
the Klein-Gordon equation for the field $\phi$ takes the form
\be
\label{waveeq}
\Bigl[ \partial_u \partial_v
- e^{(v-u)/4M}\Bigl(-{\Delta_\Omega\over r^2} + m^2 + {2M\over r^3}\Bigr)
\Bigr] r\phi(u,v,\Omeg) = 0
\ee
where $\Delta_\Omega$ denotes the scalar laplacian on the two-sphere.
For our purpose we need to consider field configurations that have a
finite frequency with respect to the Schwarzschild time $t$ and a
finite angular momentum.

\begin{center}
\leavevmode
\epsfysize=7cm
\epsfbox{figu2.eps}
\end{center}
\noindent
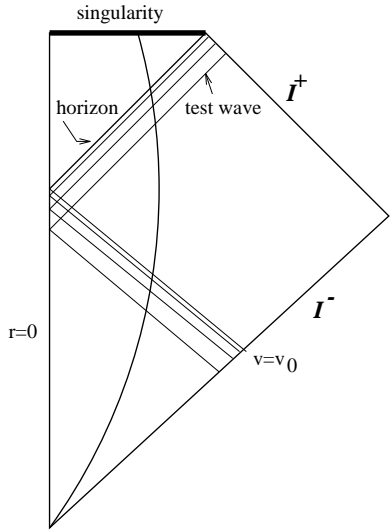
{\small Fig 2. The Penrose diagram for a black
hole that is formed through spherical gravitational collapse.
The additional lines near the horizon and $v=v_0$ indicate the light-rays
of a test-wave, that is used to determine the form of the outgoing state
of the scalar field.} 


\medskip

Since an out-going wave $\phi_{out}$ with a finite
frequency at ${\cal I}^+$ oscillates extremely rapidly near the
horizon (see figure 2), one may 
apply the geometric optics approximation to derive the form of the
incoming wave $\phi_{in}$ in the asymptotic past \cite{hawking}.
In terms of the above wave equation, this procedure becoms exact
in the region $r \rightarrow 2M$, where we have
$$
e^{ (v - u)/4M} <\!\!< 1,
$$
and thus the wave-equation (\ref{waveeq}) simplifies to
$\partial_u\partial_v \phi\!=\! 0.$ In this way we find
 that near the horizon
the field $\phi$ is decomposed into an incoming and out-going wave
\be
\phi(u,v,\Omeg) = \phi_{in}(v,\Omeg) + \phi_{out}(u,\Omeg).\label{decomp}
\ee

By matching the in- and out-signal near the region where the
horizon is formed in the initial stages of the gravitational collapse,
one then finds that these in- and out-going waves $\phi_{in}$
and $\phi_{out}$ are related via a simple reparametrization
\be
\label{reflect}
\phi_{in}(v,\Omeg) =\phi_{out}(u(v),{}^{{}_P}\Omeg),
\ee
where $\,\Omeg \equiv (\theta,\varphi)$ and $\, {}^{{}_P} \Omeg$ denotes
the corresponding anti-podal point on the two-sphere,
\ie ${}^{{}_P}\Omeg\equiv(\pi-\theta, \pi+\varphi)$.
For large $u$, the reparametrization $u(v)$ takes the
asymptotic form
\begin{equation}
\label{reflection}
u(v)  = v_0 - 4M \log \Bigl( \frac{v_0 -v}{4M} \Bigr)+const.
\end{equation}
where $v_0$ is the limiting value for the null-coordinate $v$,
describing the location of the incoming radial light-ray
that eventually coincides with the global event horizon.
The constant  on the right-hand-side
of (\ref{reflection}) depends on the details of the gravitational
collapse, and is of order $M$. In the following we will drop this
constant, because it will not be important for
our discussion. The first term $v_0$ is necessary to ensure that
our equations are invariant under time-translations which act as
simultaneous shifts of $u,v$ and $v_0$.
We note that the coordinate relation (\ref{reflection}) is
non-invertible, since it maps  the domain $v<v_0$
onto the complete range $-\infty < u <\infty$.  The equations
(\ref{reflect}) and (\ref{reflection}) play a central role in the
derivation of Hawking radiation \cite{hawking}.

To describe the quantum physics near the horizon,
the classical field variables
$\phi_{in}(v,\Omeg)$ and $\phi_{out}(u,\Omeg)$ are replaced
by second quantized field operators.
The standard free field canonical commutation relations
for the incoming fields take the form
\be
\label{comin}
\Bigl\lbrack \phi_{in}(v_1,\Omeg_1),\del_{v_2}\phi_{in}
(v_2,\Omeg_2)\Bigr\rbrack= -2\pi i \delta(v_{12})
\delta^{(2)}(\Omeg_{12}),
\ee
and a similar commutation relation holds  for the out-going fields.
The initial Hilbertspace on ${\cal I}^-$
is generated by the creation operators obtained from $\phi_{in}$
\be
\phi_{in} (v,\Omeg) = \sum_{l,m}
\int_{-\infty}^\infty\! {d\omega\over \sqrt{2\pi\omega}}\, \Bigl(
a^\dagger_{\omega lm} e^{i\omega v}+ a_{\omega lm} e^{-i\omega
v}\Bigr)Y_{lm}(\Omeg)
\ee
For each given initial state $|\psi\rangle_{in}$, we would
like to calculate the quantum state $|\psi\rangle_{fin}$
on a final Cauchy slice, which asymptotically approaches
the Cauchy surface ${\cal I}^+ \cup {\cal H}^+$, formed by
asymptotic future infinity and the event horizon.

We can expand the out-going field on ${\cal I}^+$ as
\be
\phi_{out} (u,\Omeg) =
\sum_{l,m}\int_{-\infty}^\infty\! {d\omega\over \sqrt{2\pi\omega}}\, \Bigl(
b_{\omega lm} e^{i\omega u}+ b^\dagger_{\omega lm} e^{-i\omega u}\Bigr)Y_{lm}
(\Omeg)
\ee
to obtain the creation operators that generate the $out$-Hilbert space.
Ignoring the back-reaction, these $out$-modes can be expressed in terms of the
$in$-modes, by expanding the relation (\ref{reflection}) describing the
propagation backwards in time to ${\cal I}^-$ in modes. One finds the
following linear relation
\newcommand{\xip}{{\omega'}}
\ba
\label{Bogol}
b_{\omega lm}\is
\int_0^\infty \!\! d\xip\, (\, \alpha_{\omega\xip}a_{\xip lm}+ 
\beta_{\omega\xip}a^\dagger_{\xip lm}),\nonu
b^\dagger_{\omega lm} \is\int_0^\infty \!\! d\xip\, (\,
\alpha^*_{\omega\xip}a^\dagger_{\xip lm} + \beta^*_{\omega\xip} a_{\xip lm}).
\ea
Up to some irrelevant phase, the Bogolyubov coefficients have the
asymptotic form
\ba
\label{coeff}
\alpha_{\omega\xip} \is  \,e^{-i(\xip-\omega)v_0}  \,
{e^{2\pi M \omega}\Gamma(1-i4M\omega)\over2\pi\sqrt{\omega(\xip+i\epsilon)}},
\nonu
\beta_{\omega\xip} \is \,e^{+i(\xip+\omega)v_0}
{e^{-2\pi M \omega}\Gamma(1-i4M\omega)
\over 2\pi\sqrt{\omega(\xip\!-\!i\epsilon)}}.
\ea
This relation between the asymptotic modes completely determines
the asymptotic form of the out-going state
on future infinity ${\cal I}^+$. It is described by a mixed state,
since the out-going fields on ${\cal I}^+$ cover only part of the
final Cauchy surface. We refer to Hawking's original
paper \cite{hawking} for a more detailed discussion of the properties
of this mixed state.

\newsection{Classical back-reaction.}

One of the central assumptions in the standard derivation  \cite{hawking}
of the out-going
radiation spectrum is that the incoming and out-going fields can to a very
good approximation be treated as free fields. In particular, it is
assumed that the commutator between them
\be
\Bigl[\phi_{in}(v,\Omeg_1), \phi_{out}(u,\Omeg_2)\Bigr]
\qquad
\label{commute?}
\ee
vanishes for $v>v_0$. The underlying classical intuition is that
the fields $\phi_{in}(v,\Omeg)$ with $v>v_0$ will propagate without
much disturbance into the region behind the black hole horizon, and
thus become unobservable from the out-side. However, this intuition
ignores the gravitational interactions between the in- and out-going
particles. Our aim in the following is to investigate the consequences
of these interactions for the derivation of the final state.

First let us briefly recall why in principle one could expect that
gravitational interactions can become important.
An important feature of the mapping (\ref{reflect})  is that an out-going
wave with a certain frequency $\omega$ translates into a in-wave with
infinitely many oscillations along $v=v_0$.
For example, in the $s$-wave sector we have
\be \label{s-wave}
e^{i\omega u(v)} =
e^{i\omega v_0} \Bigl( {v_0-v\over 4M}\Bigr)^{-4iM\omega}\theta(v_0-v).\ee
Hence a generic
out-wave, when propagated back in time, carries a very large out-going
stress-energy near the horizon. Although for the propagation of a
non-singular initial state this stress-energy
manifests itself only in the form of {\it virtual} fluctuations,
i.e. of purely negative frequency in a local inertial frame,
it can in principle still lead to non-trivial
gravitational effects. In order to investigate this point,
we will begin with
a description of these interactions at the classical level.

\newsubsection{Classical dynamics of the horizon.}

As a preparation, let us study the effect of a  spherical shell of matter
with energy $\delta M$ that falls into the black hole at some late
advanced time $v_1$. In particular, we want to know how this influences
the advanced time $v_0$ at which the global event horizon forms.
It is clear that due to the additional matter the Schwarzschild radius
of the black hole increases by an amount $2\delta M$, and so,
after $v=v_1$ the global event horizon coincides with the new black hole
horizon at $r=2M+2\delta M$. By tracing the corresponding
light-rays back to the origin $r=0$ we discover that the global event
horizon originates at a time $v_0+\delta v_0$ that is slightly earlier
than $v_0$ (see fig. 3). Explicitly, we find
\be
\label{deltatau}
\delta v_0=- 4 \delta M e^{-(v_1 -v_0)/4M}.
\ee
Even though this variation seems negligible,
it leads to a significant effect on out-going light-rays.
 As illustrated in figure 3, a light-ray that originally
would have reached the outside observer at some retarded time
$u$, will as a result of the shift $$v_0\ra v_0+\delta v_0,$$
arrive at a much later time $u+\delta u$.
Using  (\ref{reflection}) one easily shows that
\be
\label{deltau}
\delta u(u) = -4M\log\Bigl(1+{\delta v_0 \over 4M} e^{(u-v_0)/4M}\Bigr).
\ee
Notice that even for a very small perturbation $\delta v_0<0$
the time-delay $\delta u(u)$ becomes infinite
at a finite time $u_{lim}\! -\! v_0 \!\sim\! -\! 4M\log (|\delta v_0|/M)$.
The physical interpretation of this fact is
that a light-ray that is on its way to reach the asymptotic
observer at some time $u\! >\! u_{lim}$ will as a result of
the in-falling shell, cross the event-horizon and
be trapped inside the black-hole horizon.
It follows that an out-going wave $\phi_{out}(u,\Omeg)$ corresponding
to a given in-coming wave $\phi_{in}(v,\Omeg)$ is transformed, as a result
of the additional infalling matter,
into $$\phi_{out}\ra\phi_{out}(u+\delta u(u),\Omeg),$$
where $\delta u(u)$ is given above. This is quite a dramatic effect for a
generic out-wave.

\begin{center}
\leavevmode
\epsfysize=7.5cm
\epsfbox{figu3.eps}
\end{center}
\noindent
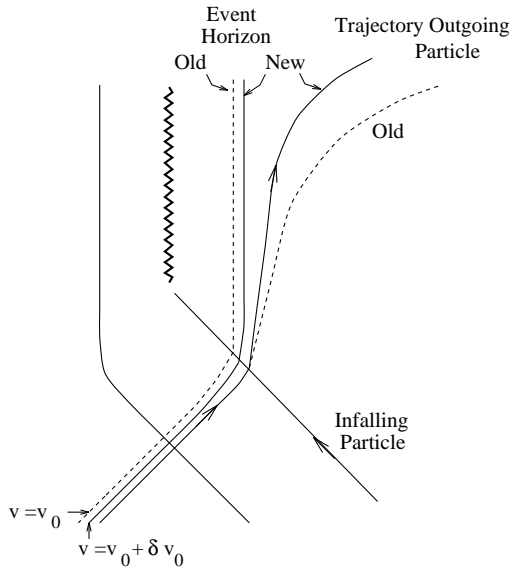
{\small Fig 3. An infalling shell of matter changes the position of the
horizon by a small amount, but due to the
redshift it has
a large effect on the trajectories of out-going light-rays.}

Let us now turn to the question of how to incorporate the gravitational
self-interactions of $\phi$ into the description of the wave-propagation.
The basic observation is that the presence of the scalar field $\phi$
also leads to small changes in the black hole mass $M$ and in the
position of the horizon due to incoming energy flux carried by the
$T_{vv}$ component of its stress-energy tensor. It is easy to show that
the stress-energy $T_{vv}\! =\! -\hf(\del_v\phi_{in})^2$ in a small
interval between $v_1$ and
$v_1+\Delta v_1$ induces a change in the mass $M$ equal to
\be
\label{deltaM}
\delta M =\int  d^2\Omeg \int_{v_1}^{\strut{}^{v_1+\Delta v_1}}\!\!\!\!\!\!\!
dv\,\,T_{vv}(v,\Omeg)
\ee
Just as in the case of an infalling matter shell, this leads to
a small correction $\delta v_0$ in the formation time $v_0$.
At first it may seem a good idea to describe this effect by substituting
(\ref{deltaM}) into (\ref{deltatau}). However, an important difference
with the previous situation is that the incoming stress-energy $T_{vv}$
is not necessarily spherically symmetric, and therefore it is reasonable
to expect that the shift $\delta v_0$ depends on the angular direction
$\Omeg$. Indeed, within a certain linearized approximation of the
Einstein equations one can derive, along the lines of \cite{dray},
 the following expression for $\delta v_0$
\ba
\label{deltav}
\delta v_0(\Omeg_1)
\is  8 \int \! d^2\Omeg_2\, f(\Omeg_1,\Omeg_2) P_{in}(\Omeg)\nonu
\label{pin}
P_{in}(\Omeg) \is  \int_{v_0}^{\infty}\! dv
\; e^{(v_0-v)/{4M}}
T_{vv} ( v , \Omeg)
\ea
where $f(\Omeg,\Omeg^\prime)$ satisfies
\begin{equation}
(\triangle_\Omega -1) f(\Omeg_1,\Omeg_2)
= - 2 \pi \delta^{(2)} ( \Omeg_{12}) .
\label{laplacian}
\end{equation}
Here $\Delta_\Omega$ denotes the Laplacian on the sphere.
The function $f(\Omeg_1,\Omeg_2)$ describes the response
of the position of the horizon at angular direction $\Omeg_1$, due to
a localized stress-energy influx from the direction $\Omeg_2$.
Notice that the expression (\ref{deltav}) contains the same exponential
factor as in (\ref{deltatau}), and thus for field configurations
with a finite energy only represents a very small correction to $v_0$.
The derivation of the result (\ref{deltav}) is summarized in the Appendix.

Let us make a short comment about the choice of the lower integration
limit at $v=v_0$. In reality this lower limit is not sharply
determined, because we should also take into account the
gravitational back-reaction due to infalling particles for $v<v_0$.
As a practical way of dealing with this technical complication,
we will in the following simply adopt as a {\it model}, that all
gravitational interactions are simply turned off below the critical
line $v=v_0$. Classically, this indeed seems a reasonable procedure,
since incoming particles at $v<v_0$ will not fall into the
black hole (provided their energy is not too large) and will therefore
not generate any shift in the critical time.

Now let us return to the problem of wave propagation on the black
hole background. As reviewed in section 2, the out-going fields are
related to the $in$-fields $\phi_{in}(v)$ for $v<v_0$
via the time-evolution from the ${\cal I}^-$ to ${\cal I}^+$.
We now propose to incorporate the
effect of the back-reaction in the relation between $\phi_{in}$ and
$\phi_{out}$ via the
substitution $v_0\!\ra\! v_0\!+\!\delta v_0(\Omeg)$ in
equation (\ref{reflection}). In this way we obtain
\begin{equation}
\phi_{out} (v , \Omeg ) = \phi_{in} (v(u) + \delta v_0(\Omega),
{}^{{}_P} \Omeg )
\label{freflection}
\end{equation}
where
\be
\label{vu}
v(u) = v_0 - 4M e^{(v_0-u)/4M},
\ee
and $\delta v_0(\Omeg)$ is given above.

\newcommand{\UU}{\,\mbox{\large $\cal U$}}


\newsection{Quantum mechanical back-reaction.}

In this section we will study how these interactions
can be incorporated at the quantum level.
In particular, we are interested in how they affect
the propagation of the $\phi$ quantum state. The idea will be to follow
the original work of Hawking \cite{hawking} as much as possible,
except that we replace the
relation (\ref{reflect}) between the in- and out-going waves by its
corrected version (\ref{freflection}).
Hence the relation between the asymptotic in and out-waves becomes
non-linear.

In the following discussion, an important role will be played
by the in- and out-going components of the stress-energy tensor.
Recall that as quantum operators, $T_{vv}$ and $T_{uu}$ not only
measure the energy flux, but are also the generators of coordinate
transformations in the $v$ and $u$ coordinates. For example,
the commutation relation of $T_{vv}$ with $\phi_{in}$ reads
\be
\label{Tvcom}
\Bigl\lbrack T_{vv} (v_1, \Omeg_1 ) , \phi_{in} \,( v_2, \Omeg_2 )
\Bigr\rbrack = 2 \pi i  \delta^{(2)} (\Omeg_{12} )
\delta ( v_{12} ) \  \partial_v \phi_{in}\, ( v, \Omeg_2 ).
\ee
and a similar relation holds between $T_{uu}$ and the out-going field
$\phi_{out}$.

\newsubsection{The algebra of $in$- and $out$-fields.}

An immediate consequence of the gravitational
back-reaction is that it invalidates the assumption that the
asymptotic in- and outgoing fields $\phi_{in}(v,\Omeg)$ at $v>v_0$
and $\phi_{out}(u,\Omega)$ can be treated as independent, commuting
variables. As we will now show, the interaction described above
implies that this commutator is in fact replaced by a non-trivial
and non-local `exchange algebra'.

We will assume that the gravitational interaction
can be incorporated in the relation between quantum operators
$\phi_{in}$ and $\phi_{out}$ via the semi-classical procedure described above,
by including the correction (\ref{deltav}) to the critical time $v_0$.
The correspondence principle guarantees that, within a reasonable
energy range, this is a valid approximation.
The diffeomorphism between the asymptotic $in$ and $out$-waves
thus depends on the quantum stress-energy tensor
$T_{vv}= -\hf (\del_v\phi_{in})^2$ contained in $\delta v_0$.
This implies that, as a quantum operator, the new critical time
$v_0+ \delta v_0(\Omeg)$
no longer commutes with the incoming fields. Using (\ref{deltav})
and (\ref{Tvcom}), we find
\be
\label{deltacom}
 \Bigl[ \delta v_0 (\Omeg_1 ), \phi_{in} (v, \Omeg_2 ) \Bigr]
= - 16 \pi i f( \Omeg_1 , \Omeg_2 ) e^{ (v_0-v)/4M}
\partial_v \phi_{in} ( v, \Omeg_2 )
\ee
It will be useful, therefore, to make the dependence of the
out-going variables on
$\delta v_0(\Omeg)$ explicit. To this end, we note that the
relation (\ref{freflection})-(\ref{vu}) can formally be inverted as follows
\footnote{The fact that $\delta v_0$ is an operator-valued quantity
in principle could introduce a problem with normal ordering at higher orders
in this expansion. In first instance, however, we will ignore
this point and simply exponentiate the linearized interaction
between the in and out-going modes. This procedure amounts to the ladder
or eikonal approximation to linearized gravity, which, in the kinematical
regime of interest, is known to provide the correct leading order result
\cite{planckscat}.}
\be
\label{eshift}
\phi_{out}\Bigl(u,\Omeg\Bigr) =
\exp\Bigl (- e^{(u-v_0)/4M}\delta v_0(\Omeg)\del_u \Bigr )
\phi_{in}\Bigl(v(u),{}^{{}_P}\Omeg\Bigr)
\ee
When we combine this relation with (\ref{deltacom}), a straightforward
calculation shows that the incoming
and outgoing fields satisfy the following exchange algebra
\be
\label{exchange}
\phi_{out} ( u , \Omeg_1 ) \phi_{in} ( v , \Omeg_2 )
= \exp\Bigl( - 16 \pi i f( \Omeg_{1},\Omeg_2 ) e^{(u - v) / 4M }
\partial_v \partial_u \Bigr)
\phi_{in} (v , \Omeg_2 ) \phi_{out} ( u , \Omeg_1 )
\ee
This result is valid for $v$ sufficiently later than $v_0$ and
for $\Omeg_1$ not too close to $\Omeg_2$.

The above formula (\ref{exchange})
is closely related to 't Hoofts two-particle $S$-matrix for Planckian
scattering \cite{planckscat} in the limit of low momentum transfer.
Note that it is
symmetric in $\phi_{in}$ and $\phi_{out}$, even though the starting point
(\ref{freflection}) seemed to be asymmetric.
A possible way to understand
this fact is that the exponentiation in (\ref{exchange}) results
from summing the contributions from multi-graviton exchange in the
eikonal approximation \cite{planckscat}.
It is further important to
note that the result (\ref{exchange}) only represents the {\it onset} of the
gravitational interaction
between the infalling matter and the out-going radiation.
The result is valid in a limited regime, because when the center
of mass energy between the $in$ and $out$-particles gets too large,
(or when the angular positions $\Omeg_1$ and $\Omeg_2$ come too close)
non-linear higher order effects will become dominant.

\newsubsection{The field operators at the horizon}

The fact that $\phi_{in}$ and $\phi_{out}$ do not commute is
physically reasonable, because the infalling matter that is in
the causal past of the operator $\phi_{out}$ in principle influences
the geometry on which the out-going field has propagated.
However, the above result not only represents
the effect of the infalling matter on the out-going
wave, but it also implies that there is a non-trivial
back-reaction effect on the infalling matter due to the presence of
the out-going fields. As the infalling wave
approaches the horizon, causality dictates that the local
field operator  $\phi$ must commute with asymptotic operators $\phi_{out}$
that describe the out-going radiation. From this we
may conclude that the in-coming field $\phi_{in}$ and the
field  at the horizon are not the same operator, but are
related via a non-trivial evolution operator.

To distinguish $\phi_{in}$
from the field at the horizon, let us denote the latter by $\phi_{hor}$.
We will try determine the proper definition of the field $\phi_{hor}$ in
terms of
$\phi_{in}$ by the condition that $\phi_{hor}$, at least formally, satisfies
\be
\Bigl[\phi_{hor}(v,\Omeg_1),\phi_{out}(u,\Omeg_2)\Bigr]=0.
\ee
The idea for finding such operators $\phi_{hor}$ is to look for
an {\it interaction representation} of the $out$-fields of the
form
\be
\label{Uphi}
\qquad \qquad
\phi_{out}(u,\Omeg) =\UU \phi_{in}(v(u),\Omeg) \UU^{-1}
\qquad \qquad v< v_0
\ee
with $v(u)$ defined in (\ref{vu}) and
where $\UU$ is some operator acting on the $in$-Hilbert space
representing the gravitational correction.
Intuitively, we may think of $\UU$ as the time-ordered
exponential of interaction Hamiltonian that describes the gravitational
self-interaction of $\phi$. Note that the relation (\ref{Uphi})
manifestly respects the canonical commutation rules  for
$\phi_{out}$ and $\phi_{in}$.

An equation of the form (\ref{Uphi}) would indeed immediately
help us in finding the field $\phi_{hor}$ near the horizon,
since one deduces from it that
the operator $\UU\phi_{in}(v,\Omeg)\UU^{-1}$  commutes with $\phi_{out}$
for $v >v_0$. This suggest that we should take
\be
\label{hor}
\qquad \qquad
\phi_{hor}(v,\Omeg) = \UU \phi_{in}(v,\Omeg)\UU^{-1} \qquad \qquad
v> v_0 \nonu
\ee
as the relation defining the fields at the horizon.

Before we determine the operator $\UU$, we like to
point out that by assuming the existence of relations of the form
(\ref{Uphi}) and (\ref{hor}), we implicitly assume
 that all infalling modes that are supported
at $v>v_0$ also fall into the black hole, and produce corresponding
horizon modes. We will see momentarily that this assumption can only safely
be made, within our approximation scheme, when we restrict to infalling
modes at sufficiently late times $v>v_0 +\Delta v$.

A useful clue that will help us find the operator $\UU$
is the apparent symmetry of the algebra (\ref{exchange})
between the incoming fields $\phi_{in}$ and out-going fields
$\phi_{out}$. We can make this symmetry more manifest by
introducing, besides the operator $P_{in}(\Omega)$ defined
in (\ref{pin}), a similar expression
\be
\label{pout}
P_{out}(\Omeg) = \int \! du\; e^{(u-v_0)/4M}T_{uu}(u,\Omeg)
\ee
in terms of the out-going fields.
By using the commutator relation
\be
\Bigl[ P_{out}(\Omeg_1),\phi_{out}(u,\Omeg_2)\Bigr] = -2\pi i
\delta^{(2)}(\Omeg_{12})e^{(u-v_0)/4M}\del_u\phi_{out}(u,\Omeg_2)
\ee
we can now replace the differential operator in the exponent of
(\ref{eshift}), and formally rewrite the
the basic relation (\ref{freflection}) between $\phi_{in}$ and $\phi_{out}$
precisely as in (\ref{Uphi}), with the operator $\UU$ given by the following
expression
\be
\label{udef}
\UU =
\exp
\Bigl[ i \int\!\! \int d\Omeg_1 d\Omeg_2
P_{out}(\Omeg_1) f(\Omeg_{1},\Omeg_2) P_{in}(\Omeg_2)\Bigr]
\ee
Inserting this result into (\ref{hor}) gives the formal
definition of the fields at the horizon.

It is instructive to work out the right-hand-side of eqn. (\ref{hor})
using the
commutation relation (\ref{Tvcom}). A simple
computation gives
\be
\label{phihor}
\phi_{hor}(v,\Omeg)=\phi_{in}(v-\Delta v(v,\Omeg),\Omeg) ,
\ee
where
\be
\label{Deltav}
\Delta v (v,\Omeg) =
4M\log\Bigl(1+e^{(v_0-v)/4M}
\int\!\! d^2\Omeg'\,  f(\Omeg,\Omeg') P_{out}(\Omeg') \, \Bigr)
\ee
A more direct derivation of equation
(\ref{phihor}) is given in the Appendix, where it is shown to represent
the effect of metric fluctuations close to the horizon.
As we see from this equation, the infalling matter fields close
to the horizon are related to the original $in$-fields via an
operator valued shift $v \rightarrow v - \Delta v$. The magnitude
of this shift is determined by the {\it outgoing} stress-energy flux
through eqn (\ref{pout}). It will become clear in the
following that, for as long as our approximation scheme
is valid, the quantum mechanical uncertainty in this quantity
$\Delta v$ will grow very fast.

A second important comment about (\ref{phihor}) is that it indicates
that our description of the gravitational interaction near the horizon
can only be trusted as long as the argument of both $\phi_{hor}$
and $\phi_{in}$ is (sufficiently) larger than $v_0$. In other
words, the definition (\ref{phihor}) of the infalling field
at the horizon requires that
\be
v > v_0 + \Delta v.
\ee
What this means is that for initial infalling field
$\phi_{in}(v,\Omeg)$  closer to $v_0$,
we can no longer say with certainty that these will reach the
black hole horizon. A more careful analysis of the interactions
in the region near $v=v_0$ will be required to determine the
fate of these fields.

\newsubsection{Back-reaction of the Hawking state.}

We will now describe how these results can be used to
investigate the effect of back-reaction on  the propagation
of a quantum state.
To this end, we return to the set-up as described at the end of
section 2, where we introduced oscillator bases for the
asymptotic $in$- and $out$-Hilbert spaces.
The $out$-modes $b_\omega$ do not generate the complete final
Hilbert space, however, and thus we will now also introduce a mode basis
of the Hilbert space near the horizon ${\cal H^+}$.
A convenient definition of modes is the following
\be
\phi_{hor} (v,\Omeg) = \sum_{l,m}
\int_{-\infty}^\infty\! {d\omega\over \sqrt{2\pi\omega}}\,
\Bigl( c_{\omega lm} e^{i\omega \tilde{u}(v)}+ c^\dagger_{\omega lm}
e^{-i\omega \tilde{u}(v)}\Bigr)Y_{lm}(\Omeg)
\ee
where
\be
\tilde{u}(v) = 4M\log(v-v_0). 
\ee
Combined together, the two sets of $b$- and $c$-modes
generate the complete final Hilbert
space, and thus the $in$-modes can be expressed in terms of them.
Using standard results as summarized in section 2 together
with the above interaction representation of the back-reaction,
one finds after a straightforward calculation (suppressing the $l,m$ labels)
\ba \UU^{-1} a_\omega\UU \is \int_0^\infty d\xip \Bigl[\,
\alpha^*_{\xip\omega} (b_\xip \!\! - \! 
e^{-4\pi M\xip} c^\dagger_\xip  \, ) -
\beta_{\xip\omega}  (\, b^\dagger_\xip\!\! - \! 
e^{4\pi M \xip}
c_\xip\, )\Bigr] \nonu
\UU^{-1} a^\dagger_\omega \UU
\is \int_0^\infty d\xip \Bigr[\, \alpha_{\xip\omega}
(\, b^\dagger_\xip  \!\! - \!  
e^{-4\pi M\xip} c_\omega\, )
- \beta^*_{\xip\omega}
(\, b_\xip  \!\! - \!  
e^{4\pi M \xip} c^\dagger_\xip\, )
\Bigr]
\ea
where $\alpha_{\omega\xip}$ and $\beta_{\omega\xip}$ are given
in (\ref{coeff}).
The form of the out-going state corresponding to the
initial vacuum state is now easily found.
The first expression for the annihilation mode $a_\omega$ shows
that final state $|\psi\rangle_{final}$ corresponding
to the $a$-vacuum $|0\rangle_{in}$ satisfies
\ba
(\, b_\omega - 
e^{-4\pi M \omega}c^{\dagger}_\omega\, )\,  \UU^{-1} \,|\psi\rangle_{final}
\is 0\nonu
(\, b^\dagger_\omega \, - 
\,  e^{4\pi M \omega} \,  c_\omega\, ) \,\UU^{-1}\, |\psi\rangle_{final} \is 0
\ea
These equations can be readily solved, if we assume that
$|\psi\rangle_{final}$
lies in the tensor product of the Fock spaces for the $b$ and
$c$-modes.  One finds
\be
\label{final}
|\psi\rangle_{final} =  \UU \, |\psi\rangle_{hawking}
\ee
where $\UU$ is the gravitational high-energy $S$-matrix
given in (\ref{udef}) and
\be
\label{abcvac}
|\psi\rangle_{hawking}
={1\over N} \exp\Bigl\lbrace  \sum_{l,m}
\int_0^\infty\!\!\!d\omega \,
e^{-4\pi M\omega} \, b^\dagger_{\omega l m} \, c^\dagger_{\omega l,-m}
\, \Bigr\rbrace \,
|0\rangle_b\otimes|0\rangle_c
\ee
where $N$ is a normalization constant.

Equation (\ref{final}) for the gravitationally corrected
final state is still rather formal. In the first place, we
need to be more specific about the integration region in the
$(u,v)$-plane that is used in the definition (\ref{udef}) of $\UU$.
The idea of our approximation procedure is to only
include the gravitational shift interaction between the $in$ and
$out$ modes very close to the horizon, and as stated before, we
simply wish to ignore all interactions near and below $v=v_0$.
Moreover we have assumed in our description that the shift interaction
extends all the way to $v=v_0$. The integration region in the
$(u,v)$-plane that we will use, therefore, is as indicated in fig 4.

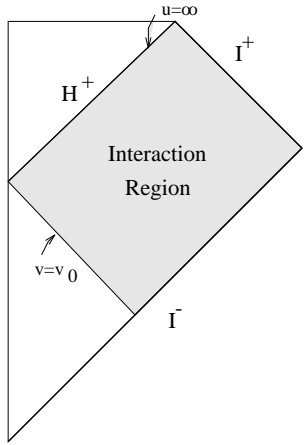
\begin{center}
\leavevmode
\epsfysize=6cm
\epsfbox{figu4.eps}
\end{center}
\noindent
{\small Fig 4. In our model we assume that the interaction region
between the in- and out-going modes is bounded by the critical line
$v=v_0$, the initial Cauchy surface at $I^-$ and the final Cauchy
surface near $H^+ \cup I^+$.}

Secondly, there are subtleties that arise from the operator nature
of the stress-energy tensors contained in $\UU$. In particular,
we need to prescribe a specific operator ordering.
The appropriate prescription here seems to use time-ordering,
as is dictated via the identification of $\UU$ as the time-ordered
exponential of the gravitational interaction Hamiltonian.
Adopting this prescription, we write $\UU$ as follows
\ba
\label{uudef}
\qquad \UU \is
{\rm T} \exp(i \int\! dt\, H_{int}) \\[2mm]
\is
{\rm T} \exp \Bigl[\, i \! \int \!\!  \int_{v_0}\!\!\!
du dv e^{(u-v)/4M} \!\! \int\!\int\! d\Omeg_1 \! d\Omeg_2 \;
T_{uu}(u,\Omeg_1) f(\Omeg_{1},\Omeg_2) T_{vv}(v,\Omeg_2)\; \Bigr]
\nonumber
\ea
where the symbol $T$ in front denotes time ordering. This formula
summarizes the leading order gravitational interaction between
the in- and out-going fields at low momentum transfer, as obtained
via the eikonal approximation.

As it stands, however, this expression still contains infinities
that arise from the singular short distance expansion of
$T_{vv}$ and $T_{uu}$ with themselves. In the following we will
assume that these infinities are regularized with the help
of some proper distance cut-off $\epsilon$.\footnote{This short distance
cut-off must regulate the short distance singularities both in the
longitudinal $(u,v)$-plane, as well as in the transverse $\Omeg$-direction.}
We will comment further on this procedure
and its physical meaning in section 4.5.

\newsubsection{Fluctuations of the Hawking state.}

We would eventually like to get some insight in the size of the
gravitational corrections we just calculated. To this end,
we first consider the magnitude of the stress-energy
fluctuations in the Hawking state, i.e. the final state before
including the back-reaction. For this purpose,
it will be convenient to choose as a basis of the final Hilbert space
the coherent state basis for the $b$- and $c$-modes
\be
\label{coherent}
|\varphi\rangle =\exp\Bigl\lbrack \int_0^\infty \!\!\! d\omega\,
(\varphi_\omega b^\dagger_\omega+
\varphi_{-\omega}c^\dagger_\omega)\Bigr \rbrack \,|0\rangle_b
\otimes |0\rangle_c .
\ee
These satisfy
the relations $ b_\omega |\varphi\rangle = \varphi_\omega|\varphi\rangle$,
$c_\omega|\varphi\rangle=\varphi_{-\omega}|\varphi\rangle$.

Using the expression (\ref{abcvac}) for the Hawking state,
one easily computes that (to simplify the
expressions we again suppress the angular dependence of the fields)
\be
\label{overlap}
\langle \psi_{hawking} | \varphi\rangle = {1\over N}
\exp[iS_0(\varphi)]
\ee
with
\be
S_0(\varphi)=\int_0^\infty\! d\omega\, e^{-4\pi M \omega}
\varphi_\omega\varphi_{-\omega}.
\ee
This result can in fact be rederived from a semi-classical
saddle-point approximation. The overlap (\ref{overlap})
is equal to the transition element ${}_{in}\langle 0|\, \varphi \, \rangle$,
evaluated in the free scalar field theory on the black hole
background. The transition element can be represented
as a functional integral over all fluctuations of the scalar field $\phi$
with specific boundary conditions at ${\cal I}^-$ and $
{\cal I}^+ \cup {\cal H}^+$  (${\cal H}^+$ denotes the event
horizon) determined by the initial state ${}_{in}\langle0|$
and final coherent
state $|\varphi\rangle$. Imposing vacuum boundary conditions
at ${\cal I}^-$ implies that $\phi$ has no
positive energy modes, while at ${\cal I}^+ \cup {\cal H}^+$ the
negative frequency part of $\phi$ is prescribed by the coherent
state $|\varphi\rangle$.

The expression $S_0(\varphi)$ can be identified with
the classical free field theory action of the saddle-point
configuration $\phi_{cl}$ associated with the coherent
state $|\varphi\rangle$:
\be
S_0(\varphi) = S_{free}(\phi^{cl})
\ee
with $\phi_{cl}$ given by the matrix element
\be
\phi_{cl} ={}_{in} \langle 0 |\phi|\varphi\rangle.
\ee
We find that the relevant classical solution associated with the
overlap (\ref{overlap}) is given by
\ba
\label{clasol}
\phi_{cl}^{in}(v)\is\int_0^\infty
\!\!d\omega\,{1\over\sqrt{\omega}} \Bigl\lbrack\varphi_\omega
\Bigl({v_0-v\over 4M}\Bigr)_+^{-4 i M\omega}+\varphi_{-\omega}
\Bigl({v-v_0\over 4M}\Bigr)^{4iM\omega}_+\Bigr\rbrack
\nonu \phi_{cl}^{out}(u)\is \int_0^\infty \!\!d\omega\,{1\over{\sqrt \omega }}
\Bigl\lbrack\varphi_\omega e^{i\omega u} + e^{-4\pi M\omega}
\varphi_{-\omega}e^{-i\omega u}\Bigr\rbrack
\ea
where the subscript $+$ indicates that the corresponding function is defined
to be analytic in the upperhalf complex $v$-plane. These configurations should
be thought of as the quantum fluctuations that are responsible for
the production of the Hawking radiation.

The magnitude of the stress-energy tensor $T_{\mu\nu}(\phi_{cl})$ depends
on the value of the parameters $\varphi_\omega$, but also on
the behavior of the modes $e^{i\omega u}$ and $(v-v_0)^{-4iM\omega}$ in the
various regions of the black-hole geometry. The
components that potentially become large are
$T_{vv}$ and $T_{uu}$.
We first consider the behaviour of the $T_{vv}$-component
associated with the in-coming field $\phi^{in}(v)$. We find
\be
\label{tvv}
T_{vv}(\phi^{cl}) = {16 M^2\over (v-v_0)^2} \int_0^\infty\! d\omega\,
\Bigl\lbrack T_\omega 
\Bigl({v_0-v\over 4M}\Bigr)_+^{-4i  M\omega}+ T_{-\omega}
\Bigr({v-v_0\over 4M}\Bigr)_+^{4iM\omega}\Bigr\rbrack
\ee
where
\be
\label{tvv2}
T_\omega =  \int_0^\omega\!\!d\omega^\prime\,
\half \sqrt{\omega^\prime(\omega-\omega^\prime)}
\varphi_{\omega^\prime}\varphi_{\omega-\omega^\prime}
+\int_0^\infty d\omega^\prime\,e^{-4\pi M\omega^\prime}
\sqrt{\omega^\prime(\omega+\omega^\prime)}
\varphi_{-\omega^\prime}\varphi_{\omega+\omega^\prime}
\ee
and a similar expression can be given for $T_{-\omega}$ ($\omega>0$).

We see that for finite non-vanishing values for the
parameters $\varphi_\omega$ the stress-energy tensor is
in general singular near $v=v_0$. The $T_{uu}$-component,
representing the out-going stress-energy flux,
is related to $T_{vv}$ by reflection off the $r=0$ boundary.
In the Kruskal coordinate $x^-= -e^{-u/ 4M}$
the out-going stress-tensor is generically also
singular near the horizon as soon as the $\varphi_\omega$ differ
by a small amount from their exact expectation value.
It is easy to convince oneself that
such fluctuations are also really present:
The average magnitude
of $\varphi_\omega$ in the Hawking state $|\psi\rangle_{hawking}$ is
\be
\overline{\varphi_{\omega} \varphi^\ast_{\omega'}} =
\langle\ n_\omega +1\rangle \delta(\omega-\omega^\prime)
\ee
and this indicates that the fluctuations of the
(absolute value)${}^2$ of $\varphi_\omega$ are comparable to the
fluctuations in the particle number density observed in the out-state.
Thus for generic coherent out-states the parameters $\varphi_\omega$
will indeed differ by a finite amount from their average value, and this
will result in large, super-Planckian stress-energy fluctuations
near the horizon.

\medskip

\newsubsection{Gravitational corrections and complementarity.}

We will now comment on the expression
(\ref{final})-(\ref{uudef}) for the corrected
final state. The following remarks will be mostly qualitative,
as some of the relevant calculations are left for a future publication.

Let us first slowly turn on the back-reaction. Using the expressions
(\ref{final}) and (\ref{uudef}), we obtain the leading order
correction to the final state eqn (\ref{overlap}), by adding
to the classical free field action $S_{free}(\phi_{cl})$ an
interaction term
\be
\label{overlap2}
\langle \psi_{final} | \varphi\rangle = {1\over N}
\exp\Bigl[i S_0(\varphi) + iS_{int}(\varphi)\Bigr]
\ee
with $S_{int} = \int \! dt \, H_{int}$ as given in eqn. (\ref{uudef}).
In leading order, this interaction term can
be evaluated on the unperturbed classical field configuration $\phi_{cl}$.

How large is this leading order correction? This turns out to
be a somewhat subtle question. In principle, one can
explicitly compute this correction term by inserting
(\ref{tvv})-(\ref{tvv2}) in our expression (\ref{uudef})
for $S_{int}$. One then finds, however, that the magnitude
of the corrections critically depends on how one treats the
end-points of the integration over the $u$ and $v$ coordinates.
If one defines the integration region, as indicated in fig 4, to
have infinitely sharp boundaries at the horizon $u=\infty$ and
the critical line $v=v_0$, one will find that the corrections to
the final state on ${\cal H}^+ \cup {\cal I}^+$ in fact become very large.
(This can easily be seen from the above expressions for $T_{vv}$.)
However, if instead one cuts off the interactions
at some finite distance from the horizon, as measured
in some local coordinate system, one will find that the corrections are
bounded, and, because of the exponentially growing redshift, will eventually
get smaller with time for the final state on ${\cal I}^+$.

A similar cut-off dependence arises,  if one tries to
compute higher order corrections by further expanding the exponent
in (\ref{uudef}). In this case, additional short distance singularities
arise because of the time ordering prescription. In principle,
in a finite consistent theory of gravity, such as perhaps string theory,
these singularities should be smoothed out by the short distance
gravitational dynamics. However, given our lack of understanding of this
short distance theory, it seems a reasonable procedure to represent
its effect by introducing some cut-off scale $\epsilon(x)$.

To understand this procedure somewhat better, let us recall the discussion
of section 1.1  on the space-time complementarity principle.
There we also introduced a cut-off scale $\epsilon(x)$. The point of that
discussion was that a reasonable gravitational cut-off must not only
regulate the integrals of $T_{\mu\nu}$, but must also act at
a more fundamental level and truncate the Hilbert space by
eliminating all states with wavelengths smaller than the cut-off
scale. Indeed, the short-distance singularity in the operator
product relation between two stress-tensor operators arises due to
the contribution of intermediate states with arbitrarily short
wave-lengths. Regulating the singularity in the OPE's of $T_{\mu\nu}$
is therefore equivalent to throwing out these singular states.

What does this cut-off procedure imply for the calculation of the
gravitational corrections to the Hawking spectrum?
Suppose we take for $\epsilon$ some proper
distance cut-off in the region very close to the horizon.
This will indeed ensure that we can find a
reasonable and controlled answer for the form of the final state
on ${\cal H}^+$. However, we immediately run into trouble
if at the same time we want to calculate the form of the
out-going state on ${\cal I}^+$, because practically all states
on ${\cal I}^+$ correspond to singular states near the horizon that
were thrown out by our cut-off procedure.
A proper distance cut-off near the horizon is therefore not suitable
for computing the out-going spectrum (see  eg. \cite{jacobson}).

Instead, to calculate the state on ${\cal I}^+$, we are forced to
choose a cut-off that allows the Hilbert space to contain very high
frequency out-going modes near the horizon, since the corresponding quantum
fluctuations were used in the zeroth order calculation.
The price one pays, however, is that these modes
interact violently with all in-falling particles, via the
stress-energy fluctuations computed in section 4.4.
This will in turn result in a fast growing quantum uncertainty
in the geometry near the horizon, which is most clearly
exhibited by inserting the expression for $T_{uu}$, as obtained from
(\ref{tvv})-(\ref{tvv2}), into equation (\ref{phihor}).
If we nevertheless insist on producing a reliable
semi-classical description of the entire final state on
${\cal H}^+ \cup {\cal I}^+$, we are forced to truncate the
Hilbert space of the matter near the horizon
so that it contains only very low frequency incoming modes.
In this Hilbert space we can reasonably trust
the calculations of the out-going state, and we find that
the corrections are finite. Qualitatively the out-going state
will look very much like the thermal state as derived by Hawking.
However, it is clear that due the necessary truncation of the
Hilbert space, the entropy associated with the black hole is
drastically reduced compared to the conventional free field
result.

Summarizing, we thus
propose that the calculation of the out-going state
can be performed in a controlled
fashion by working in a theory with such a truncated
Hilbert-space, with the idea that this procedure represents a reasonable
effective description of some underlying
consistent theory of quantum gravity.
By regulating the calculation in this way, one produces a finite result
for the gravitationally corrected final state, which however
critically depends on the choice of cut-off.

\newsection{Summary}

We have presented a new method for computing the gravitational
corrections to the emission spectrum of an evaporating black hole.
In this procedure we included gravitational interactions
between the matter fields as they propagate from and
initial Cauchy slice near ${\cal I}^-$ to a final slice near
${\cal H}^+ \cup {\cal I}^+$. Contrary to common expectations,
we find that the gravitational interactions that play a role
in this calculation are not suppressed, and can lead to potentially
large physical effects.

It is clear that some parts of our approximation scheme can be
improved. In particular, it should be in principle be possible
to exactly compute the leading order correction to the final state,
that results from single graviton exchange between the in-
and out-going virtual particles. This will in particular
clarify some issues of our calculation related to the
interactions near $v=v_0$.

In going to higher orders, conventional field theoretical methods
will become inadequate, because gravity is non-renormalizable.
Thus in any approach based on local field theory, one is forced
to introduce some effective description by introducing a cut-off.
The physical picture that emerges from these considerations is
described in sections 1.1. and 4.5.

\medskip

\medskip

\noindent
{\bf Acknowledgements.}

\noindent
The research of E.V. and of H.V. is partly supported by an
Alfred P. Sloan Fellowship, and by a Fellowship of the Royal Dutch
Academy of Sciences. The research of Y.K. and H.V.
is supported by NSF Grant PHY90-21984
and the Packard Foundation. H.V. also acknowledges the
support by a Pionier Fellowship of NWO.

\renewcommand{\thesection}{A}
\renewcommand{\thesubsection}{A.\arabic{subsection}}

\vspace{15mm}
\pagebreak[3]
\setcounter{section}{1}
\setcounter{equation}{0}
\setcounter{subsection}{0}
\setcounter{footnote}{0}

\begin{flushleft}
{\large\bf Appendix: Derivation of (\ref{freflection}) and (\ref{phihor}).}
\end{flushleft}

We start from the assumption that the back-reaction effects are small,
and therefore
we allow ourselves to work in the weak field
approximation. Weak gravitational fields are represented
by small perturbations around the classical metric
\be
ds^2  = ds^2_{cl} + h_{\mu\nu}dx^\mu dx^\nu.
\ee
In first approximation $h_{\mu\nu}$ satisfies the linearized Einstein equations
with source equal to  the stress-tensor $T_{\mu\nu}$ of the scalar field.
Hence, $h_{\mu\nu}$ may be expressed as
\be
\label{hsol}
h_{\mu\nu}(x) = 8 \G \int \! d^4 y \, D_{\mu\nu}{}^{\lambda\sigma}(x,y)
T_{\lambda\sigma}(y),
\ee
where $D_{\mu\nu}^{\lambda\sigma}(x,y)$ denotes the
propagator for the graviton field on the classical black hole
background. Next one should substitute this perturbation back
into the scalar wave-equation to obtain the corrected
equation that to this order includes all the gravitational self-interactions.

The calculation can be considerably simplified by focusing
only on the interactions between the in-coming and out-going waves
near the horizon, which turn out to be the most significant.
It turns out to be convenient to work in the Kruskal parametrization
in terms of $x^\pm$. We want to find an approximate solution
of the linearized Einstein equations, that is valid close to the horizon
and takes into account the back-reaction effect of the
$T_{\pm\pm}$ components of the stress-energy tensor. Thus
we assume that $T_{++}$
and $T_{--}$
are separately conserved: \ie
$\del_{\mp} T_{\pm\pm}=0,$
and that the other components of $T_{\mu\nu}$ are negligible.
We further take the following {\it Ansatz} for the metric near the black hole
horizon
\be
\label{qruskal}
ds^2 =
- {32 M^3\over r} e^{-r/2M}(dx^+ + h_{--}(x^-,\Omeg) dx^-) (dx^- +
h_{++}(x^+,\Omeg)dx^+)
+ r^2 d^2 \Omeg
\ee
The linearized Einstein equations for this metric are
\be
\kappa (\triangle_\Omega-1) h_{\pm\pm}
= T_{\pm\pm}
\ee
with $\kappa = \frac{2^9 GM^4}{e}$. This equation integrates to
\be
\label{sol}
h_{\pm\pm}(x^\pm,\Omeg) = {1\over \kappa}
\int d\tilde{\Omeg} \; f(\Omeg,\tilde{\Omeg})
T_{\pm\pm}(x^\pm,\tilde{\Omeg}).
\ee
Next we consider  the wave equation for
the scalar field $\phi$ in the background
metric (\ref{qruskal}). Just as in section 2, one finds that near the horizon
it takes  a simplified form:
\be
\nabla_+\nabla_-
 \phi=0,\qquad\qquad \quad\nabla_\pm = \del_\pm -h_{\pm\pm}\del_{\mp}.
\ee
Using the fact that $\lbrack \nabla_+,\nabla_-\rbrack =0$, one sees that
the classical solutions again separate into a sum of an incoming and
an outgoing part: $\phi=\phi_{in}+\phi_{out}$, with $\nabla_+\phi_{out}=0$
and $\nabla_-\phi_{in}=0$. Solving these first order differential equations
for $\phi_{in}$ and $\phi_{out}$, one finds that
the outgoing field $\phi_{out}$  must be of the form
\be
\label{effout}
\phi_{out}(x^-,x^+,\Omeg) =
\phi_{out}\bigl(x^- +\int_{x^+}^\infty dy^+ h_{++}(y^+,\Omeg),\Omeg\bigr)
\ee
Here we have chosen our integration limits such that for $x+\ra\infty$
there is no shift in the argument. The matching with the $in$-field
occurs near the point $x^+_0= e^{v_0/4M}$.\footnote{Note
that we are indeed only considering
the interaction that take place in the horizon region $x^+ > e^{v_0/4M}$.
For values of $x^+$ near the critical time the dynamics is rather more
complicated.}
Inserting the expression (\ref{sol}) for $h_{++}$,
taking $x^+\ra \infty$, and translating the result back to the $u,v$
coordinates produces the result (\ref{freflection}).

Similarly, the infalling matter is also sensitive to the metric fluctuations
induced by the out-going matter.
The equation of motion for infalling waves reads
\be
\partial_- \phi_{in}(x^+,x^-,\Omeg) = h_{--}(x^+,\Omeg)
\partial_+\phi_{in}(x^+,x^-,\Omeg)
\ee
and can in a similar way  be formally solved via
\be
\label{infall}
\phi_{in}(x^+,x^-,\Omeg) =
\phi_{in}\bigl(x^+-\int^{\infty}_{x^-}dy^-h_{--}(x^+,\Omeg),\Omeg\bigr)
\ee
Substituting $x^-=0$ and re-expressing the result in $u,v$ coordinates
leads to equation (\ref{phihor}).
Finally, we want to repeat that the solutions
of the linearized Einstein equation and scalar wave-equation that we just
described
are valid only in the close neighbourhood of the black hole horizon.
We have ignored all other components of $T_{\mu\nu}$ except $T_{\pm\pm}$, and
therefore strictly speaking our analysis applies only to classical fields
$\phi$ for which these other components are not too large.

\end{document}